\documentclass[conference, letterpaper]{IEEEtran}
\IEEEoverridecommandlockouts
\usepackage[utf8]{inputenc}
\usepackage[T1]{fontenc}

\usepackage{eulervm} 
\usepackage{dsfont} 

\usepackage[american]{babel}

\usepackage{amsmath}
\usepackage{amsthm}

\usepackage{amssymb}
\usepackage{wasysym}
\usepackage{textcomp}

\usepackage{tikz}
\usetikzlibrary{shapes, patterns, positioning}

\usepackage[hidelinks]{hyperref}

\usepackage[
	sort&compress,
	nameinlink,
	noabbrev,
        capitalise
]{cleveref}

\usepackage[acronym]{glossaries-extra}
\setabbreviationstyle[acronym]{long-short}

\usepackage{csquotes}

\usepackage{cite}

\usepackage{multirow}
\usepackage{graphicx}
\usepackage{xcolor}
\usepackage{soul}

\usepackage{booktabs}
\usepackage{flushend}
\usepackage[
    protrusion=true,
    activate={true,nocompatibility},
    final,
    tracking=true,
    kerning=true,
    spacing=true,
    factor=1100]{microtype}
\SetTracking{encoding={*}, shape=sc}{40}

\usepackage{mathtools}

\usepackage{braket}
\usepackage{sansmath}

\usepackage{ifdraft}
\usepackage{etoolbox, xpatch}

\usepackage[normalem]{ulem}
\usepackage{tabularx}

\usepackage{makecell}
\usepackage{textcomp}
\def\nobreakbefore{%
  \relax\ifvmode\else
    \ifhmode
      \ifdim\lastskip > 0pt\relax
        \unskip\nobreakspace
      \fi
    \fi
  \fi
}
\let\oldcite\cite
\renewcommand\cite{\nobreakbefore\oldcite}
\usepackage{enumitem}

\usepackage{siunitx}               

\DeclareSIUnit\px{px}

\theoremstyle{plain}

\theoremstyle{definition}

\newacronym{gcd}{GCD}{Greatest Common Divisor}

\usepackage{caption}
\begin{document}

\title{
Optimal Partitioning of Quantum Circuits using Gate Cuts and Wire Cuts
}

\author{
\thanks{This work was partially funded by the Carl Zeiss foundation.}

Sebastian Brandhofer,$^{1, 2}$ Ilia Polian,$^{1}$ Kevin Krsulich$^{2}$ 
\\\\
\begin{minipage}[c]{\textwidth}
\centering
\small $^1$ Institute of Computer Architecture and Computer Engineering and Center for Integrated Quantum Science and Technology, University of~Stuttgart, Stuttgart, Germany, e-mail: \{sebastian.brandhofer, ilia.polian\}@iti.uni-stuttgart.de\\
\small $^2$ IBM Quantum, IBM TJ Watson Research Center, Yorktown Heights, NY, USA, e-mail: kevin.krsulich@us.ibm.com
\end{minipage}
}

\maketitle

\begin{abstract}
A limited number of qubits, high error rates, and limited qubit connectivity are major challenges for effective near-term quantum computations.
Quantum circuit partitioning divides a quantum computation into a set of computations that include smaller-scale quantum (sub)circuits and classical postprocessing steps.
These quantum subcircuits require fewer qubits, incur a smaller effort for satisfying qubit connectivity requirements, and typically incur less error.
Thus, quantum circuit partitioning has the potential to enable quantum computations that would otherwise only be available on more matured hardware.
However, partitioning quantum circuits generally incurs an exponential increase in quantum computing runtime by repeatedly executing quantum subcircuits.
Previous work results in non-optimal subcircuit executions hereby limiting the scope of quantum circuit partitioning.

In this work, we develop an optimal partitioning method based on recent advances in quantum circuit knitting.
By considering wire cuts and gate cuts in conjunction with ancilla qubit insertions and classical communication, the developed method can determine a minimal cost quantum circuit partitioning.
Compared to previous work, we demonstrate the developed method to reduce the overhead in quantum computing time by 73\% on average for 56\% of evaluated quantum circuits.
Given a one hour runtime budget on a typical near-term quantum computer, the developed method could reduce the qubit requirement of the evaluated quantum circuits by 40\% on average.
These results highlight the ability of the developed method to extend the computational reach of near-term quantum computers by reducing the qubit requirement at a lower increase in quantum circuit executions.
\end{abstract}

\maketitle

\section{Introduction}
The ability of near-term quantum computers to run effective quantum computations is hindered by an insufficient number of qubits, high error rates, and restrictive connectivity constraints \cite{future_bravyi, preskill-nisq, swap_overhead}.
Quantum circuit partitioning can help with these obstacles by enabling to compute parts of a quantum computation separately, either sequentially on a single quantum computer or in parallel on multiple individual quantum computers \cite{multi_cz, wire_cutting_no_ancilla2, circuit_knitting, priv3, priv16, harrows, optimal_wire_cutting, wire_cutting_no_ancilla0, wire_cutting_no_ancilla1, priv1, priv4, cutqc, cutqc2, wirecutting0}.
Hereby, the requirements on the connectivity and the number of qubits for the partial quantum computations are reduced, which typically leads to less errors and an extended computational reach \cite{cutqc, cutqc2, wirecutting0}.
For instance, an otherwise intractable quantum computation (or: quantum circuit) could be performed by first applying quantum circuit partitioning to yield smaller quantum (sub)computations that each \emph{can} be performed on a near-term quantum computer.

However, the benefits of quantum circuit partitioning are offset against an exponential runtime increase on the quantum computer.
This increase in runtime may prevent the application of partitioning or diminishes potential quantum advantages.
It is therefore crucial to rigorously minimize the overhead in quantum computing time for quantum circuit partitioning.
Specifically, partitioning a quantum circuit requires to:
\begin{enumerate}
    \item Determine wire or gate dependencies within a quantum circuit that prevent a suitable partitioning.
    \item Prepare a partitioning by resolving the dependencies using circuit cutting or knitting techniques \cite{circuit_knitting, priv16, priv3, harrows, wire_cutting_no_ancilla0, wire_cutting_no_ancilla1, wire_cutting_no_ancilla2, optimal_wire_cutting}.
    \item Compute (or: sample from) the determined partitioning on a quantum computer, i.e. execute the individual subcircuits included in the partitions.
    \item Recombine the results of the subcircuit computations to yield the result of the partitioned quantum circuit.
\end{enumerate}
The last step of quantum circuit partitioning can be performed in polynomial time for a class of recent dependency resolution techniques called \emph{quantum circuit knitting} \cite{circuit_knitting, optimal_wire_cutting}.
The execution of quantum subcircuits in the third step of quantum circuit partitioning requires the same steps as for unpartitioned quantum computations.
In addition, recent advances \cite{circuit_knitting, optimal_wire_cutting, wire_cutting_no_ancilla0, wire_cutting_no_ancilla1} resulted in optimal dependency resolution techniques (step two).
These optimal dependency resolution techniques have not been considered in previous work \cite{cutqc, cutqc2, wirecutting0, harrows} for the first step of partitioning and thus they can not determine a minimal cost partitioning.

The work at hand utilizes the dependency resolution techniques introduced in \cite{circuit_knitting, optimal_wire_cutting} to address the first step of quantum circuit partitioning by:
\begin{itemize}
    \item Extending \cite{optimal_wire_cutting} by introducing a novel construction for single wire dependency resolution at minimal cost without requiring ancilla qubits.
    \item Developing a satisfiability modulo theories (SMT) model that enables an optimal partitioning of quantum circuits 

    \item Evaluating the developed SMT model on quantum circuits amenable to partitioning in principle \cite{revlib, maslov, qaoa, vqe_he, ghz}.
\end{itemize}

The remainder of this work is organized as follows.
In \cref{knit sec:cuts} quantum circuit partitioning, gate cuts and wire cuts are introduced.
\cref{knit sec:circ_knitting} describes dependency resolution via quantum circuit knitting \cite{circuit_knitting}, and the novel construction for single wire dependency resolution.
In \cref{knit sec:related} the overhead of existing dependency resolution methods is described and previous quantum circuit partitioning methods \cite{cutqc, harrows, wirecutting0} are compared to this work in detail.
Then, in \cref{knit sec:method} the developed SMT model and its preprocessing steps are described whereas \cref{knit sec:results} presents the results of the evaluation and \cref{knit sec:conclusion} concludes the work.

\section{Quantum Circuit Partitioning with Gate Cuts and Wire Cuts}\label{knit sec:cuts}
Dependencies given by quantum gates or qubit wires within a quantum circuit enforce that the quantum circuit is executed on one \emph{partition}, i.e. on a set of connected qubits that can interact with each other potentially after inserting auxiliary operations \cite{olsq, ibmqx_mapping, sabre, wille_teleport, mapping_teleportation, rydberg}.
Here, a \emph{gate dependency} represents the requirement of qubits involved in the same multi-qubit gate to be connected.
A \emph{wire dependency} represents the continuity of a qubit assignment in time, i.e. the state corresponding to a qubit can not be transferred to another qubit between the execution of quantum gates.
Without partitioning, all qubits affected by wire or gate dependencies must belong to the same set of connected qubits.
A \emph{wire cut} resolves a wire dependency and a \emph{gate cut} resolves a gate dependency.

\begin{figure}[t!]
  \centering
  \includegraphics[width=0.80\linewidth]{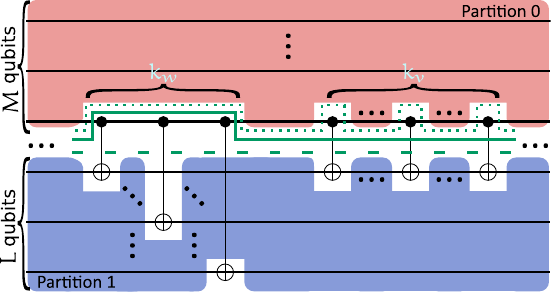}
  \caption{Example quantum circuit partitioning using gate cuts (dashed green line), wire cuts (dotted green line), and both (solid green line). The red (top) and blue (bottom) colored blocks indicate a high density of two-qubit gates where partitioning incurs a prohibitive cost in general. \label{knit fig:cuts}    
}
\end{figure}

\cref{knit fig:cuts} demonstrates an example partitioning of a quantum circuit into two partitions while resolving dependencies within the quantum circuit by using different combinations of cuts (green lines).
The target quantum circuit contains $k_w$ CNOT quantum gates arranged as demonstrated in the left part of the quantum circuit and $k_v$ CNOT quantum gates arranged in the ladder structure on the right side of the quantum circuit.
The red (top) and blue (bottom) colored blocks indicate parts of the quantum circuit with a high density of two-qubit gates, i.e. where performing a partitioning incurs a prohibitive cost.
Partitioning using only gate cuts is depicted by a dashed green line, only wire cuts are depicted by a dotted green line, and both types of cuts are depicted by a solid green line in \cref{knit fig:cuts}.

As will be explained in the following sections, the result of the quantum circuit partitioning using any of the indicated cuts is two separate partitions:
the entangling gates between those partitions have been replaced by quantum operations local to the respective partition and classical postprocessing.
Depending on the selected cut types, the number of required cuts varies from $2+2k_v$ for only wire cuts and $k_{w} + k_v$ for only gate cuts to $\min(2, k_{w})+k_v$ for the combined usage of gate cuts and wire cuts.
Any selection of these cuts reduces the qubit requirement on a quantum computer, i.e. instead of requiring $L+M$ qubits for the depicted quantum circuit, only $\max(L, M)$ qubits are required on a quantum computer.

While the number of cuts is always minimal when employing both types of cuts, using only wire cuts can be preferable to using only gate cuts (and vice versa) depending on the exact structure of the quantum circuit, i.e. the number of CNOT quantum gates $k_w$ and $k_v$ in this example quantum circuit.
Previous work on quantum circuit partitioning only employed wire cuts to yield a quantum circuit partitioning \cite{cutqc, harrows, wirecutting0}.
Note that the cost of a cut -- while still exponential -- can vary significantly depending on the cut location, the availability of ancilla qubits, classical communication, and the used dependency resolution technique \cite{circuit_knitting, priv1, priv4, priv3, priv16, harrows, cutqc, cutqc2, wirecutting0, wire_cutting_no_ancilla2, multi_cz, optimal_wire_cutting, wire_cutting_no_ancilla0, wire_cutting_no_ancilla1, wire_cutting_no_ancilla2}.

\section{Gate Cuts and Wire Cuts via Quantum Circuit Knitting}\label{knit sec:circuit_knitting}\label{knit sec:circ_knitting}

The goal of partitioning is to assign qubits and quantum gates to different smaller partitions that then can be executed separately from each other.
This will in general violate gate and/or wire dependencies, and methods such as circuit knitting aim at overcoming the effects of such violations \cite{circuit_knitting, priv1, priv4, priv3, priv16, harrows, cutqc, cutqc2, wirecutting0, wire_cutting_no_ancilla2, multi_cz, optimal_wire_cutting, wire_cutting_no_ancilla0, wire_cutting_no_ancilla1, wire_cutting_no_ancilla2}.
By resolving these dependencies, a (connected) quantum circuit can be divided into separate partitions.
These partitions contain \emph{quantum subcircuits} whose individual computation results (samples) are combined to estimate the result of the unpartitioned quantum circuit computation.
Quantum subcircuits can be computed separately from each other, i.e. quantum gates do not occur between quantum subcircuits, and typically require fewer qubits and time individually for computation than the unpartitioned quantum circuit.
Furthermore, these quantum subcircuits are at most connected to each other via classical communication \cite{circuit_knitting, optimal_wire_cutting, dynamic_circuits}.
The number of times these subcircuits must be executed, i.e. the \emph{sampling overhead}, grows exponentially for every resolved dependency and defines the total quantum runtime of the partitioned quantum circuit.
Thus, minimizing the sampling overhead is a crucial objective for quantum circuit partitioning.

\begin{figure}[t!]
  \centering
  \includegraphics[width=1\linewidth]{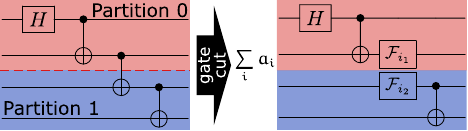}
  \caption{Dividing a 4-qubit GHZ state into two partitions using quantum circuit knitting to resolve gate dependencies. \label{knit fig:gate_cut}
}
\end{figure}

In quantum circuit knitting (see \cref{knit fig:gate_cut}) \cite{circuit_knitting, priv16, priv3}, a gate is cut by representing the unitary channel $\mathcal{U}$ corresponding to a quantum gate $g$ by a quasiprobability decomposition (QPD)
\begin{equation}\label{knit eq:qpd}
    \mathcal{U} = \sum_{i} a_{i} \mathcal{F}_{i},
\end{equation}
where $\mathcal{F}_{i}$ are operations that can be realized locally (including potential classical communication) and weights $a_{i}$ indicate how to estimate the computation of $\mathcal{U}$.
The quantum gate $g$ is randomly replaced by operations $\mathcal{F}_{i}$ with a probability that depends on $a_{i}$, each time a partition is computed that includes the quantum gate $g$.
One such replacement is called a subcircuit and the incurred sampling overhead, i.e. the increase in the number of subcircuit executions, to estimate the output of the original quantum circuit scales as $\kappa^{2}$ with the $\kappa$-factor $\kappa := \sum_{i} |a_{i}|$ for each cut gate.
There may be multiple QPDs for a quantum gate each with their individual $\kappa$\emph{-factor} \cite{circuit_knitting}.
The smallest number $\kappa$ of a gate according to \cref{knit eq:qpd} is called the $\gamma$\emph{-factor} in this work. 
In general, the more entangling a quantum gate is, as quantified by the robustness of entanglement \cite{entanglement_robustness, circuit_knitting}, the higher the minimum incurred sampling overhead of quantum circuit knitting will be.

After determining all subcircuits introduced by circuit knitting, the expectation value of the operator measurement on the original quantum circuit is estimated by a simple sum of the measurement outcomes of the subcircuits weighted by their corresponding $a_i$ \cite{circuit_knitting}.

Dependency resolution techniques are able to reduce the sampling overhead of multiple, \emph{simultaneous cuts} by using ancilla qubits and classical communication \cite{circuit_knitting, wire_cutting_no_ancilla0, wire_cutting_no_ancilla1}.
Specifically for the quantum circuit knitting technique used in this work, multiple simultaneous gate cuts are realized via quantum gate teleportation protocols \cite{circuit_knitting}.
These protocols can realize a quantum gate across two partitions by consuming a preexisting Bell state shared between these partitions in addition to operations local to these partitions.
The sampling overhead of cutting simultaneous Bell state generations is lower than individually generating and cutting these Bell states \cite{circuit_knitting} (see \cref{knit sec:related} for details).
However, generating these Bell states requires one additional ancilla qubit on each of their respective partitions, and the quantum gate teleportation protocols require classical communication between partitions.

\subsection*{Wire Dependency Resolution via Gate Cuts}
\begin{figure}[htbp]
  \centering
  \includegraphics[width=1\linewidth]{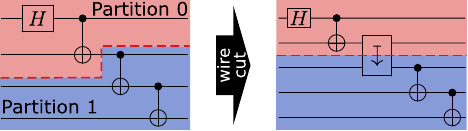}
  \caption{Dividing a 4-qubit GHZ state into two partitions by resolving a wire dependency via gate cuts. \label{knit fig:wire_via_gate} 
}
\end{figure}

\cref{knit fig:wire_via_gate} shows how to transform a wire dependency to a gate dependency that can then be cut by regular quantum circuit knitting (see previous \cref{knit fig:gate_cut}) \cite{circuit_knitting}.
At the location of the wire cut, a set of quantum gates (indicated by a boxed vertical arrow) is inserted that transfers the state left to the wire cut to a new qubit that is used in subsequent operations.
A gate cut is then applied to the inserted set of quantum gates in a subsequent step, which removes the newly introduced gate dependencies.
Thus, the qubit storing the quantum state left to the wire cut and the new qubit to which the state was transferred can be part of different partitions.
Note that in general, the sum of overall used qubits increases by one for each wire cut in the absence of qubit reuse techniques \cite{reset_hua, reuse} (see the right of \cref{knit fig:wire_via_gate}).

\begin{figure}[htbp]
  \centering
  \includegraphics[width=0.44\linewidth]{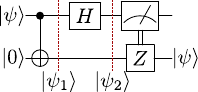}
  \caption{'Move' circuit to relocate the state $\ket{\psi}$ from the top qubit to the bottom qubit without further ancilla qubits; $\ket{\psi_{1}}$ and $\ket{\psi_{2}}$ are the states corresponding to \cref{knit eq:psi1} and \cref{knit eq:psi2}, respectively. By inserting this circuit, a wire cut can be reduced to gate cutting the only occuring CNOT gate.
  \label{knit fig:move_circuit}    
}
\end{figure}

There are several options to implement the operation given by the boxed vertical arrow at the location of the wire cut: e.g. a swap gate, a quantum teleportation circuit \cite{optimal_wire_cutting}, or the quantum circuit described in \cref{knit fig:move_circuit}, which we call a 'move' circuit.
All of these options incur different costs in terms of ancilla qubits, sampling overhead, and the requirement for classical communication. While the 'move' circuit and the quantum teleportation circuit incur the lowest possible sampling overhead (see \cref{knit sec:related}), they also require the ability to perform classical communication between separate partitions.
Furthermore, the quantum teleportation circuit requires one extra ancilla qubit.

We now show that the 'move' circuit indeed moves the state $\ket{\psi} = \alpha\ket{0} + \beta\ket{1}$ of a qubit to an ancilla qubit in the $\ket{0}$ state. 
The 'move' circuit consists of a Bell measurement and subsequent phase correction \cite{qc10th}.
After the first CNOT gate in the 'move' quantum circuit, the two-qubit state $\ket{\psi}\ket{0}$ becomes 
\begin{equation}\label{knit eq:psi1}
    \ket{\psi_{1}} = \alpha\ket{00}+\beta\ket{11}.
\end{equation}
Applying the Hadamard gate to the top qubit leads to 
\begin{equation}\label{knit eq:psi2}
    \ket{\psi_{2}} = \alpha\left(\ket{0}+\ket{1}\right)\ket{0}+ \beta\left(\ket{0}-\ket{1}\right)\ket{1}.
\end{equation}
The measurement on the top qubit either yields a zero or a one with equal probability. In case of a 'zero' measurement, the state of the bottom qubit is:
\begin{equation}
    \bra{0}\cdot \ket{\psi_{2}} = \alpha (0 + 1)\ket{0} + \beta (0+1)\ket{1} = \alpha\ket{0} + \beta\ket{1},
\end{equation}
and in the case of a 'one' measurement on the top qubit, the state of the bottom qubit is
\begin{equation}
    \bra{1}\cdot \ket{\psi_{2}} = \alpha (0 + 1)\ket{0} + \beta (0-1)\ket{1} = \alpha\ket{0} - \beta\ket{1}.
\end{equation}
Thus, correcting the phase, i.e. applying a Z-gate, on the bottom qubit depending on the measurement result of the top qubit, yields the state $\ket{\psi}$ on the bottom qubit.

\section{Related Work}\label{knit sec:related}
In this section, the overhead of gate and wire dependency resolution is summarized.
Furthermore, the overhead of previous automatic quantum circuit partitioning methods \cite{cutqc, harrows, cutqc2} is compared to this work.
Dependency resolution techniques were developed in multiple works \cite{circuit_knitting, priv1, priv4, priv3, priv16, harrows, cutqc, cutqc2, wirecutting0, wire_cutting_no_ancilla2, multi_cz, optimal_wire_cutting, wire_cutting_no_ancilla0, wire_cutting_no_ancilla1, wire_cutting_no_ancilla2}.
These works can be distinguished by their application to gate cuts \cite{multi_cz, priv1, priv4, priv16} or wire cuts \cite{harrows,  priv3, optimal_wire_cutting, wire_cutting_no_ancilla0, wire_cutting_no_ancilla1, wire_cutting_no_ancilla2, cutqc, cutqc2, wirecutting0}.
In addition, the work in \cite{priv3, priv16, harrows, cutqc, cutqc2, wirecutting0} assumes no classical communication (CC) while the work in \cite{circuit_knitting, optimal_wire_cutting, wire_cutting_no_ancilla2, wire_cutting_no_ancilla1} reduces the overhead of dependency resolution by inserting ancilla qubits and using CC between individual partitions.

\cref{knit tab related} gives an overview of the minimal sampling overhead of dependency resolution techniques.
The sampling overhead varies depending on the type of applied cut (gate cut or wire cut) and the availability of CC as well as ancilla qubit insertions.
A single individual wire cut incurs a multiplicative sampling overhead of 9 (e.g. this work and \cite{optimal_wire_cutting, wire_cutting_no_ancilla1, wire_cutting_no_ancilla0}) when CC and ancilla qubits are available and 16 otherwise \cite{harrows, cutqc, cutqc2, optimal_wire_cutting, wire_cutting_no_ancilla0, wire_cutting_no_ancilla1, wire_cutting_no_ancilla2}.
In contrast to the work in \cite{optimal_wire_cutting}, recent approaches in \cite{wire_cutting_no_ancilla0, wire_cutting_no_ancilla1} and this work achieve the minimal sampling overhead for a \emph{single} wire cut without requiring additional ancilla qubits.
With CC and ancilla qubit insertions a minimal sampling overhead of $(2^{k+1}-1)^2$ is required for $k$ arbitrary wire cuts \cite{circuit_knitting, wire_cutting_no_ancilla0, wire_cutting_no_ancilla1} and $16^k$ otherwise.

The sampling overhead of cutting CNOT quantum gates, swap gates, and the conditional-rotation (CR) quantum gate is given on the right side of \cref{knit tab related}.
Without CC, the multiplicative sampling overhead is $9^k$ \cite{priv3}, $49^k$ \cite{priv16}, and $(1+2|\sin(\theta)|)^{2k}$ \cite{circuit_knitting} respectively for cutting $k$ CNOT, swap and conditional-rotation quantum gates. While CC and additional ancilla qubits allow the reduction of sampling overhead to $(2^{k+1}-1)^2$, $16^k$, and to less than $4^k$ for cutting $k$ CNOT, swap, and conditional-rotation quantum gates respectively \cite{circuit_knitting}.
As determining a QPD and the corresponding $\gamma$-factor for an arbitrary quantum gate is computationally hard, the minimal sampling overhead of other classes of quantum gates is largely unknown \cite{circuit_knitting}.
\subsection*{Related Work on Quantum Circuit Partitioning}
Previous quantum circuit partitioning methods resolve dependencies via wire cuts only and without CC \cite{cutqc, cutqc2, harrows}.
As seen in the previous \cref{knit tab related}, wire cuts incur a multiplicative sampling overhead of $16^k$ for individual and simultaneous cuts without CC and ancilla qubits and $\mathcal{O}(4^{k})$ otherwise.
Depending on the quantum circuit structure (see \cref{knit fig:cuts}), these works may also additionally incur a larger number of cuts than partitioning methods using gate cuts and wire cuts in conjunction.
As gate cuts are not considered in previous partitioning works \cite{cutqc, cutqc2, harrows}, minimal cost partitionings based on cutting gates with a low sampling overhead such as CR gates can not be determined.
Thus, previous quantum circuit partitioning methods incur a larger number of cuts at a higher cost in sampling overhead depending on the given quantum circuit.

In addition to a worse sampling overhead, previous work on quantum circuit partitioning also incurs an exponential overhead in classical postprocessing \cite{cutqc, cutqc2, wirecutting0} whereas this work is based on quantum circuit knitting with an efficient classical postprocessing \cite{circuit_knitting}.

\begin{table}[t!]
\addtolength{\tabcolsep}{-0.45em}
\caption{Minimal sampling overhead of cutting depending on cut type, ancilla qubits and classical communication (CC).\label{knit tab related}}
\begin{tabular}{@{}lccccc@{}}
                                                             & \multicolumn{2}{c}{Wire Cuts} & \multicolumn{3}{c}{k Gate Cuts}                    \\ \cmidrule(lr){2-3} \cmidrule(lr){4-6}
                                                             & Single    & k Arbitrary       & CNOT            & swap   & CR$(\theta)$            \\ \midrule
\begin{tabular}[c]{@{}l@{}}CC and\\ Ancilla\end{tabular}     & 9         & $(2^{k+1}-1)^2$   & $(2^{k+1}-1)^2$ & $16^k$ & $(\leq 4)^k$            \\ 
\begin{tabular}[c]{@{}l@{}}No CC, \\ no Ancilla\end{tabular} & 16        & $16^k$            & $9^k$           & $49^k$ & $(1+2|\sin(\theta)|)^{2k}$ \\ \bottomrule

\end{tabular}
\end{table}

\subsection*{Number of Available Cuts for Quantum Circuit Partitioning}
\begin{figure}[b!]
  \centering
  \includegraphics[width=0.9\linewidth]{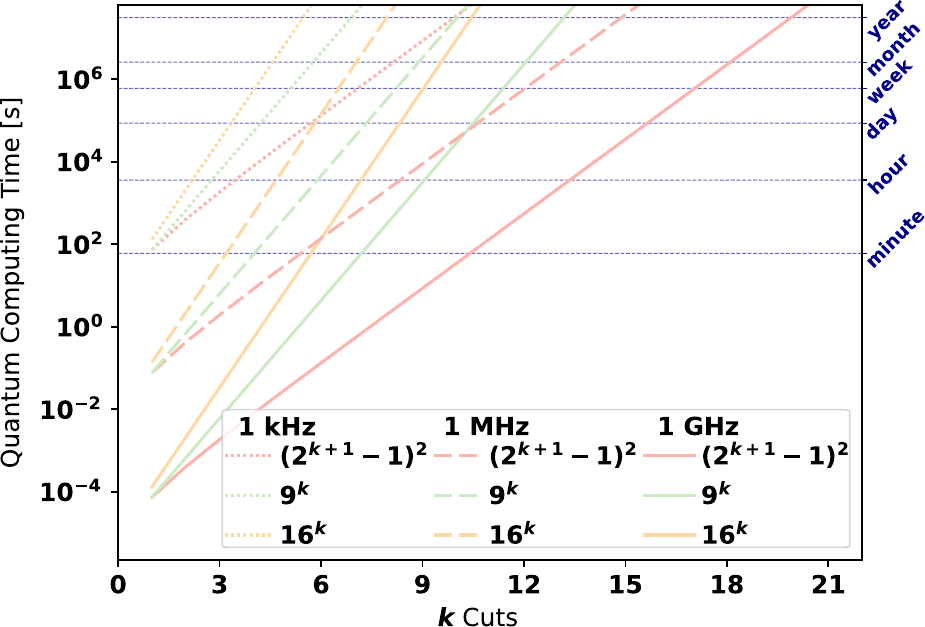}
  \caption{Overhead in quantum computing time for $k$ cuts that incur a multiplicative sampling overhead of $(2^{k+1}-1)^2$, $9^k$ and $16^k$ at different quantum computer sampling frequencies. \label{knit fig:exp_scaling}
}
\end{figure}

In general, the number of applicable cuts is limited by the incurred exponential sampling overhead. 
\cref{knit fig:exp_scaling} shows the runtime of a quantum circuit computation on the y-axis for a number of cuts on the x-axis at different sampling frequencies and multiplicative sampling overheads assuming eight thousand measurements as a basis.
The depicted sampling overheads exemplarily represent the cost of cutting CNOT $k$ gates with CC and ancilla qubits ($(2^{k+1}-1)^2$) as well as cutting single wires with CC and without ancilla qubits at $9^k$ (e.g. this work).
Furthermore, the cost of cutting single wires without CC $(16^k)$ as employed in previous quantum circuit partitioning methods \cite{cutqc, cutqc2, harrows} is also depicted.
Note that quantum gates such as conditional-rotation gates may incur a sampling overhead smaller than $\mathcal{O}(4^{k})$ depending on the applicable rotation angles \cite{circuit_knitting}.

For a total available runtime of one day for the entire quantum execution, quantum computers with a sampling frequency of 1 kHz, 1 MHz, and 1 GHz would be able to accommodate 5, 10, and 15 cuts, respectively at the minimal sampling overhead per CNOT cut.
However, at a sampling overhead of $9^k$ ($16^k$) per cut, only 4, 7, 10 $(3, 5, 8)$ cuts can be accommodated at a sampling frequency of 1 kHz, 1 MHz, and 1 GHz.
The repetition rate of current quantum computers based on superconducting qubits is in the order of roughly 1 kHz to 20 kHz \cite{weber, ibmq}.

The limited number of cuts available at even large quantum computing budgets warrant a runtime-intensive and rigorous quantum circuit partitioning approach that can determine a suitable partitioning with provably minimal sampling overhead for a given quantum circuit.

\section{Optimized Partitioning of Quantum Circuits}\label{knit sec:method}
\begin{figure}[t!]
  \centering
  \includegraphics[width=1.0\linewidth]{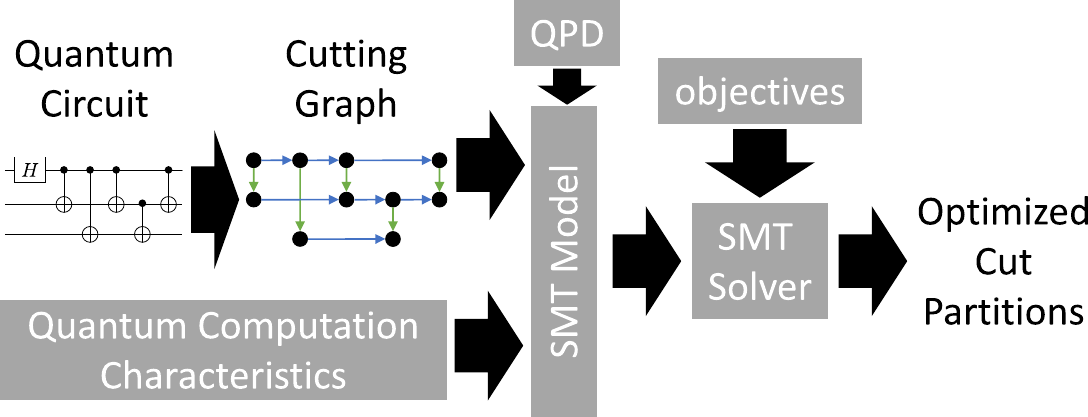}
  \caption{Individual steps of the developed quantum circuit partitioning method. \label{knit fig:model_flow}    
}
\end{figure}

In this section, we describe the developed satisfiability modulo theories model (SMT) in detail, outlining the necessary preprocessing steps, model variables, model constraints, and evaluated objective functions.
The individual steps are shown in \cref{knit fig:model_flow}.
First, the target quantum circuit that should be partitioned is input to a preprocessing step that compiles the quantum circuit to two-qubit quantum gates and that generates the cutting graph used as a basis for the developed SMT model.
Then, the generated graph is input to the SMT model together with a description of the QPDs including their $\gamma$-factor as well as a specification of the available quantum computing resources in terms of qubits, computing time, and sampling frequency.
In the last step, an SMT solver, such as the Z3 solver \cite{z3solver}, is used to solve the input SMT model subject to defined objective functions, i.e. finds an assignment to the model variables that is optimal with respect to the defined objective function.
The assignment to the model variables can then be used to generate the individual subcircuits using the individual QPD of the chosen cuts as described in \cref{knit sec:circ_knitting}.

\subsection{Preprocessing Steps} \label{knit sec:preproc}
\begin{figure}[htbp]
  \centering
  \includegraphics[width=1.0\linewidth]{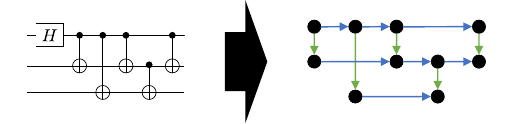}
  \caption{The cutting graph (right) of an example input quantum circuit (left). Horizontal edges (blue) represent potential wire cuts and vertical edges (green) represent potential gate cuts. \label{knit fig:graph}    
}
\end{figure}

First, the target quantum circuit is compiled to two-qubit gates \cite{qiskit} and then transformed to the \emph{cutting graph} $\mathcal{G}=(V, G \cup W)$ where $W$ represents the set of edges corresponding to wire cuts and $G$ represents the set of edges corresponding to quantum gate cuts, by the steps detailed in the following section.
An example of this preprocessing step applied to an example quantum circuit is shown in \cref{knit fig:graph} with wire cut edges $W$ colored in blue and quantum gate cut edges $G$ colored in green.
First, single-qubit gates are removed from the input quantum circuit and each two-qubit quantum gate $g$ between qubits $g_u$ and $g_v$ is represented by two vertices $g_u$ and $g_v$ in graph $\mathcal{G}$ and an edge $s_{g}=(g_{u}, g_{v}) \in G$.
Then, for each pair of two-qubit gates $(h, k)$ where $k$ is an immediate successor of $h$, an edge is added to the edge set $W$ for each overlapping qubit.
Specifically, an edge $(a, b)\in W$ is added for each vertex $a\in s_{h}$ and vertex $b\in s_{k}$ with $q_{a}=q_{b}$, where $q_{a}$, $q_{b}$ denote the qubits corresponding to vertices $a, b$.

\subsection{Model Variables}
Given the cutting graph $\mathcal{G}=(V, G \cup W)$ with $E:=(G \cup W)$ determined by the preprocessing step and a set of partitions $P$, the developed SMT model has variables:
\begin{itemize}
\item The set $O=\{o_{v, p}\;|\; v\in V, p \in P\}$ - where $o_{v, p}$ evaluates to true if the gate qubit vertex $v$ is assigned to partition $p$.
\item The set $C=\{c_{e}\;|\; e\in E\}$ - where $c_{e}$ evaluates to true if the quantum circuit is cut at the edge $e$.
\item The set $B=\{b_{e}\;|\; e\in E\}$ - where $b_{e}$ evaluates to true if the wire or gate cut at edge $e$ is cut simultaneously, i.e. requires ancilla qubits on the individual partitions.
\end{itemize}
The maximum number of partitions $|P|$ can be upper-bounded by the number of vertices in the cutting graph as then every vertex would be assigned a different partition.
\subsection{Model Constraints}
The variables $o_{v, p}\in O$ define the assignment of gate qubit vertices to partitions, and thus the exact cut positions, whenever the assignment mismatches between two vertices of an edge $e\in E$.
Thus, a cut variable $c_{e=(u, v)}$ is determined by
\begin{equation}
	c_{e=(u, v)} := \bigvee_{p\in P} o_{u, p} \neq o_{v, p},
\end{equation}
while ensuring that a vertex $v\in V$ is assigned to exactly one partition.
This is enforced by ensuring that no vertex is assigned twice
\begin{equation}
    o_{v, p} \rightarrow \neg o_{v, p'} \qquad  p, p' \in P\text{ with } p\neq p'
\end{equation}
and that at least one partition is assigned to each vertex $v$:
\begin{equation}
    \bigvee_{p\in P} o_{v, p}.
\end{equation}

The selection of simultaneous cuts is encoded using variables $b_{e}$ that imply $c_{e}$. 
The assignment to variables $o_{v, p}\in O$ defines the number of partitions
and together with the assignments to cut variables $c_{e}$ and $b_{e}$ also define the number of qubits $Q_p$ in a partition $p$:
\begin{equation}
\begin{split}
Q_{p} := \sum_{v\in I\subseteq V} o_{v, p} + \sum_{e=(u, v)\in W} c_{e} \wedge o_{v, p} \\
+ \sum_{e=(u, v)\in E}b_{e}\wedge (o_{v, p} \vee o_{u, p}),
\end{split}
\end{equation}
where $I$ is the set of vertices without incoming edges from set $W$, i.e. the first vertex on each qubit.

In addition to the above constraints, a set of constraints is added that fixes the domain of the model variables such as enforcing the number of qubits in a partition to be a natural number less than the maximum allowed number of qubits in a partition, constraining the maximum number of cuts, partitions or samples.

Note that a partitioning solution where only wire cuts are employed can be obtained by enforcing cut variables $c_{g}$ with $g\in G \subseteq E$ to evaluate to false (this is the basis for evaluation in \cref{knit sec:results}).
\subsection{Objective Functions}
In the subsequent section, two objective functions are evaluated in various settings.
One objective is set to minimize the incurred increase in samples, while the second objective minimizes the qubit requirement of the target quantum circuit.
The increase in samples is given by 
\begin{equation}\label{knit eq:num_samples}
S:= \prod_{e\in E} \gamma^{2}_{e}(c_{e}\wedge \neg b_{e}) \prod_{e\in E}\gamma^{2}_{e^{(k)}} b_{e},
\end{equation}
where $\gamma_{e}$ is the individual cut $\gamma$-factor of the cut associated with edge $e\in E$ and $\gamma_{e^{(k)}}$ is the $k$ simultaneous cut $\gamma$-factor of the cut associated with edge $e\in E$.
The $\gamma$-factors used in this work are introduced by this work and the work in \cite{circuit_knitting} (see \cref{knit tab related} in \cref{knit sec:related} for an overview).
Minimizing $S$ over the satisfiable assignments to the SMT model variables then determines partitions with a minimal sampling overhead.
Note that \cref{knit eq:num_samples} can be linearized by introducing auxiliary variables and by applying the logarithm, yielding a sum instead of a product.
The maximum number of qubits in a partition can be minimized by
\begin{equation}
\min Q, \text{with } Q\geq Q_{p} \qquad \forall p\in P.
\end{equation}

\section{Results}\label{knit sec:results}
In this section, three aspects of the developed quantum circuit partitioning method are evaluated.
First, the relationship of the sampling overhead to an increasingly tight qubit requirement is quantified.
A tighter qubit requirement leads to smaller partitions and thus also to a decreasing number of qubits available on the target quantum computer.
Second, the impact of the quantum computing budget on the possible relaxation of qubit requirement is investigated to quantify the potential of applying quantum circuit partitioning in the near term.
We then conclude this section by demonstrating a reduction of sampling overhead for the developed method when compared to the sampling overhead of previous quantum circuit partitioning work \cite{cutqc, cutqc2, harrows}

The developed SMT model was evaluated with a timeout of one hour on quantum circuits with up to 40 qubits and depths of up to 51 quantum gates.
The evaluated quantum circuits consist of arithmetic functions \cite{revlib, maslov}, GHZ state preparation \cite{ghz}, hardware-efficient ansatz circuits \cite{vqe_he}, and QAOA quantum circuits for solving the maximum cut problem on random \emph{connected} graphs with additional random edges whose number amount to 10\%, 30\%, 50\% and 100\% of the number of vertices in the graph.
The evaluated quantum circuits were compiled into two-qubit gates using Qiskit\cite{qiskit}.
If not noted otherwise, the budget for the quantum computation was set to a runtime of one day on a sampling frequency of 1 MHz, which equals roughly $10^{11}$ samples.

\subsection{Sampling Overhead for a Decreasing Number of Qubits}
\begin{figure}[t!]
  \centering
  \includegraphics[width=0.9\linewidth]{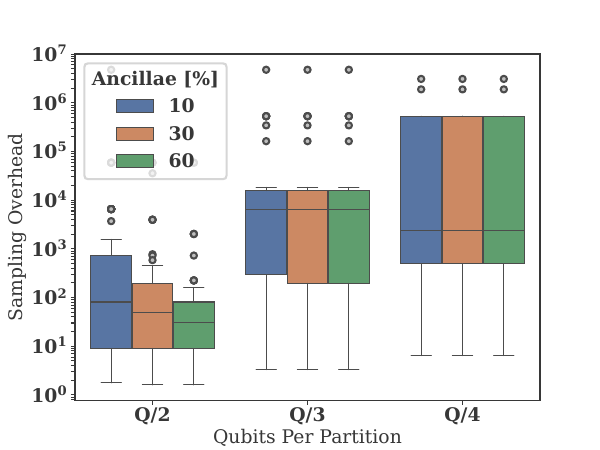}
  \caption{Sampling overhead for partitions with at most $Q/2,Q/3,Q/4$ qubits and an additional $10\%-60\%$ ancilla qubits where $Q$ is the number of qubits in the target quantum circuit. \label{knit fig:decr_qubits}
  }
\end{figure}

\cref{knit fig:decr_qubits} shows the overhead in the number of samples when imposing increasingly strict limits on the number of qubits available per partition compared to the number of qubits in the target quantum circuit.
More precisely, the number of qubits in a partition is reduced by a factor of 2, 3, and 4 while allowing for additional 10\%, 30\%, or 60\% ancilla qubits on that partition.
The additional qubits on a partition allow for further sampling overhead reduction and to utilize wire cuts for partitioning quantum circuits that would otherwise require all qubits on that partition.
For instance, a 4-qubit GHZ state quantum circuit can only be divided into two 2-qubit partitions using gate cuts.
Employing a wire cut would require one partition to have at least three qubits due to the inherently required ancilla qubit (see previous \cref{knit fig:wire_via_gate}).
Note that reducing the number of qubits in a partition by a factor of $2, 3$, and $4$ compared to the original number of qubits in the target quantum circuits implies that the quantum circuit is divided into at least $2, 3$, or $4$ partitions.

Partitioning the evaluated quantum circuits without ancilla qubits incurs the largest sampling overhead for all of the investigated partition sizes.
The sampling overhead is lowest for the first group of boxplots in \cref{knit fig:decr_qubits} where two or more partitions are allowed at most half of the qubits required in the target quantum circuit.
In this setting, the sampling overhead reduction due to simultaneous cuts is the largest.
The benefit of simultaneous cuts diminishes with a larger number of partitions the target quantum circuit is divided into and with a smaller number of ancilla qubits per partition.
The sampling overhead is largest for the third group of boxplots where each partition is allowed at most a quarter of the qubits required in the target quantum circuit.
An increase in ancilla qubits per partition does not reduce the sampling overhead further for these cases.

These results show that the developed approach is able to significantly reduce the qubit requirement of the evaluated quantum circuits in the given quantum computing budget.
Furthermore, simultaneous cuts for quantum circuit partitioning is demonstrated to be an effective tool in extending the scope of quantum circuit partitioning.

\subsection{Reduction in Qubits for Increasing Computing Budget}
\begin{figure}[t!]
  \centering
  \includegraphics[width=0.71\linewidth]{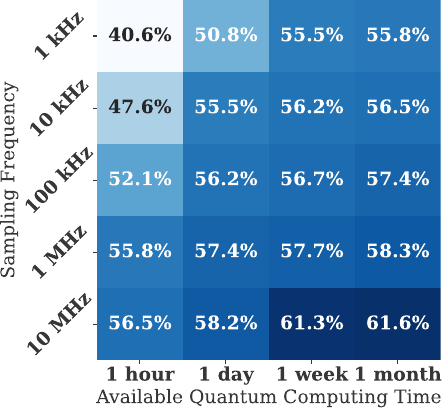}
  \caption{Reduction in partition qubit requirement for varying quantum computing runtimes.\label{knit fig:varying_budget}
}
\end{figure}

\cref{knit fig:varying_budget} shows the average reduction in qubit requirement for a computing budget varying from a sampling frequency of 1 kHz to 10 MHz and a computing time of one hour to one month.
The reduction in qubits is shown in each cell of the heatmap.
It ranges from $40.6\%$ for the lowest investigated quantum computing runtime budget of one hour and a sampling frequency of 1 kHz to $61.6\%$ for the largest investigated runtime budget of one month and a sampling frequency of 10 MHz.
These results show that even a limited quantum computing budget is sufficient to significantly reduce the qubit requirement of the evaluated quantum circuits, while the largest reduction in qubit number requires large quantum computing budgets.
Thus, quantum circuit partitioning can potentially double the size of quantum computations available on near-term quantum computers that run at sampling frequencies as low as 1 kHz at the cost of a significantly increased quantum runtime.

\subsection{Advantage of Combining Gate Cuts and Wire Cuts }
In this section, the sampling overhead of only employing wire cuts is compared to this work for dividing the evaluated quantum circuits into two partitions that each have half as many qubits as the given unpartitioned circuit.
Previous work on quantum circuit partitioning uses only wire cuts without classical communication \cite{cutqc, cutqc2, wirecutting0}.

\begin{figure}[t!]
  \centering
  \includegraphics[width=0.9\linewidth]{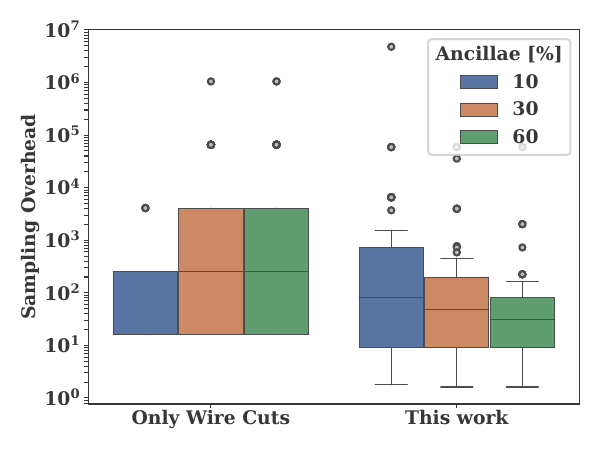}
  \caption{Sampling overhead of partitioning when using only wire cuts with each incurring $16^k$ more samples (previous work \cite{cutqc, cutqc2, wirecutting0, harrows}), and when using a combination of gate cuts and wire cuts with $10\%-60\%$ ancilla qubits per partition and classical communication.\label{knit fig:cuknit_compare}    
}
\end{figure}

\cref{knit fig:cuknit_compare} shows the sampling overhead of previous work (left) compared to the method developed in this work (right) for 10\%, 30\%, and 60\% additional qubits per partition (see \cref{knit sec:related} for a theoretical comparison).
By employing gate cuts and wire cuts as well as considering ancilla qubits and classical communication between partitions, the developed method is able to significantly reduce the sampling overhead.
In general, for roughly 56\% of the evaluated quantum circuits, the sampling overhead could be reduced by $73\%$ on average over previous work that only uses wire cuts \cite{cutqc, cutqc2, wirecutting0}.

Furthermore, ancilla qubits are often required for extending the scope of quantum circuit partitioning.
Since one ancilla qubit is required per wire cut in the absence of qubit reuse \cite{reuse, reuse_hua}, quantum circuits that require a larger number of cuts for partitioning only become tractable with a larger portion of ancilla qubits per partition.
For this work, a larger number of ancilla qubits also reduces the sampling overhead through simultaneous cuts, allowing for quantum circuit partitioning that incur a prohibitive sampling overhead otherwise.
More precisely, 41\% of the evaluated quantum circuits can not be partitioned using only wire cuts with 10\% ancilla qubits within the given quantum computing budget of one day at 1 MHz.
This reduces to 14\% when 30\% ancilla qubits are available and further to 8\% when 60\% ancilla qubits are available.
The combined use of gate cuts and wire cuts requires much fewer ancilla qubits to partition the evaluated quantum circuits within the given quantum computing budget: only 3\% of quantum circuits could not be partitioned with 10\% ancilla qubits, which reduces to 2\% for 30\% ancilla qubits or more.
Furthermore, while a larger portion of the evaluated quantum circuits could be partitioned by the method developed in this work, the partitionings also incurred a lower sampling overhead.

These results highlight the benefit of using a combination of gate cuts and wire cuts in conjunction with classical communication and ancilla qubits for quantum circuit partitioning compared to only employing wire cuts without classical communication as in previous works \cite{cutqc, cutqc2, wirecutting0}.

Note that the results in \cref{knit fig:cuknit_compare} were obtained by applying the approach developed in this work while enforcing specific values for the cut variables corresponding to gate cuts and wire cuts.
Thus, gate cuts and classical communication (for supporting simultaneous cuts) have been disabled to allow for comparison with the capabilities of previous partitioning work not based on quantum circuit knitting \cite{cutqc, cutqc2, wirecutting0}.
The previous work on quantum circuit partitioning was not directly invoked \cite{cutqc, cutqc2, wirecutting0} as they primarily minimize the overhead in classical processing instead of minimizing only the sampling overhead \cite{cutqc}.
Nevertheless, the minimal sampling overhead of $16^k$ for $k$ wire cuts was used as a basis for comparison \cite{optimal_wire_cutting}.

\section{Conclusion}\label{knit sec:conclusion}

In this work, we presented the first rigorous quantum circuit partitioning method based on quantum circuit knitting that combines wire cuts and gate cuts to soften qubit-number, error, and latency limitations of today's and near-term quantum computers.
The developed method also considers ancilla qubit insertions, the availability of classical communication, and individual gate cut costs to reduce the incurred sampling overhead further.
We introduced a 'move' quantum circuit that can be used to resolve wire cuts using quantum circuit knitting with minimal overhead and no cost in ancilla qubit.

The evaluation of the developed partitioning method demonstrated a 73\% reduction in sampling overhead on average for the 56\% of the evaluated quantum circuits over previous work that only uses wire cuts \cite{cutqc, cutqc2, harrows}.
Simultaneously, the developed method required less ancilla qubits and was able to partition a larger portion of the evaluated quantum circuits within the given qubit and runtime budget.
Even using a relatively small quantum computing budget of one hour at a sampling frequency of 1 kHz allowed reducing the required number of qubits by $40\%$. 
A reduction of $60\%$ was observed to require a quantum computing budget of 1 month at a sampling frequency of 10 MHz.
This highlights the applicability of quantum circuit partitioning at a low quantum runtime budgets.

In the future, we plan to extend the developed method by considering further quantum circuit characteristics such as the mapping effort and duration when selecting cuts for quantum circuit partitioning.

\section{Acknowledgments}
We thank Christophe Piveteau, Daniel Bhatti, Stefan Woerner and Antonio Mezzacapo for helpful discussions.
This material is based upon work supported by the U.S. Department of Energy, Office of Science, National Quantum Information Science Research Centers, Co-design Center for Quantum Advantage (C2QA) under contract number DE-SC0012704.
This work was partially funded by the Carl Zeiss foundation.

\scriptsize
\bibliographystyle{IEEEtran}
\bibliography{bibliography}

\end{document}